\documentclass[conference]{IEEEtran}\IEEEoverridecommandlockouts
\usepackage{amsfonts}
\usepackage{cite,graphicx,amsmath,amsthm}
\usepackage{subcaption}
\usepackage{caption}
\usepackage{fancyhdr}
\usepackage{dsfont}
\usepackage{array,color}
\usepackage{bm}
\usepackage{float}
\usepackage{algorithm}
\usepackage{algpseudocode}
\usepackage{multirow}
\usepackage{booktabs}
\usepackage{multirow}
\usepackage{makecell}
\usepackage{tikz}
\usepackage{hyperref}
\usetikzlibrary{positioning} 
\usepackage{xcolor} 
\allowdisplaybreaks[4]

\begin{document}
	\title{\huge Integrated Robotics Networks with Co-optimization of Drone Placement and Air-Ground Communications}

	\author{Menghao Hu$^*$, Tong Zhang$^*$, Shuai Wang$^\dagger$, Guoliang Li$^\dagger$, Yingyang Chen$^*$, Qiang Li$^*$,  Gaojie Chen$^\&$\\

$^*$ Department of  Electronic Engineering, Jinan University, Guangzhou 510632, China \\
	
$^\dagger$ Shenzhen Institute of Advanced Technology, Chinese Academy of Sciences, Shenzhen 518055, China\\
	
$^\&$ 5GIC \& 6GIC, Institute for Communication Systems (ICS), University of Surrey, Guildford GU2 7XH, UK\\
	
	\textsc{Email}: Joshua@stu2020.jnu.edu.cn

	\thanks{This work was supported by the Shenzhen Science and Technology Program under Grant RCBS20200714114956153 and Fundamental Research Funds for the
		Central Universities under Grant 11623336.}
	\vspace{-0.2in}
}

\maketitle	
	 
\begin{abstract}
	Terrestrial robots, i.e., unmanned ground vehicles (UGVs), and aerial robots, i.e., unmanned aerial vehicles (UAVs), operate in separate spaces. To exploit their complementary features (e.g., fields of views, communication links, computing capabilities), a promising paradigm termed integrated robotics network therefore emerges, which provides  communications for cooperative UAVs-UGVs applications. However, how to efficiently deploy UAVs and schedule the UAVs-UGVs connections according to different UGV tasks become challenging. In this paper, we consider the sum-rate maximization problem, where UGVs plan their trajectories autonomously and are dynamically associated with UAVs according to their planned trajectories. Although this problem is a NP-hard mixed integer program, a fast polynomial time algorithm using alternating gradient descent and penalty-based binary relaxation, is devised. Simulation results demonstrate the effectiveness of the proposed algorithm.
\end{abstract}

	
\section{Introduction}	
	
The sixth generation (6G) cellular networks aim at providing ubiquitous and high-speed communications. As such, the communications can happen in terrestrial robots, i.e., unmanned ground vehicles (UGVs), aerial robots, i.e., unmanned aerial vehicles (UAVs), and between them. Furthermore, we have witnessed many successful applications of UGVs and UAVs \cite{Ruihua,Jinchuan,9573459}. The UAVs have advantages of low communication blockage probability and high mobility, while often far away from the sensors/targets at the ground, leading to low precision of observation. On the other hand, UGVs have advantages of approaching the sensors/targets at the ground, while subject to blockages at the ground and low mobility. To exploit their complementary features (e.g.,
fields of views, communication links, computing capabilities),
a promising paradigm termed integrated robotics network
emerges, providing communications for cooperative UAVs-UGVs applications.


The integration of UAVs and UGVs has garnered significant attention due to its immense potential for various applications \cite{niu2022unmanned,Lingxiao,Xiangqian,Jie,Jianqiang,Jianq,Zhao}. In a recent study \cite{niu2022unmanned}, a vision-based UAV-assisted path planning for multiple UGVs was proposed, highlighting the benefits of integrating UAVs and UGVs in a collaborative manner. Another study was presented in \cite{Lingxiao}, where the authors introduced an aerial-terrestrial vehicle equipped with a binding mechanism capable of connecting with both aerial and terrestrial UGVs.
Incorporating millimeter-wave (mmWave) communication into the UAVs-UGVs cooperative system was explored in \cite{Xiangqian} as a means to enhance communication capabilities between the vehicles. The study investigated the advantages of integrating mmWave communication in facilitating efficient cooperation between UAVs and UGVs. The authors of \cite{Jie} proposed a taxonomy for classification of existing UAVs and UGVs to provide a tool and guidance for analyzing
various UAVs–UGVs coordination patterns. 
Addressing the challenges of environment mapping and path planning in UAVs-UGVs cooperative systems, \cite{Jianqiang} conducted research to develop effective strategies for building environment maps and planning paths.  
A specific application of the UAVs-UGVs cooperative system for illegal urban building detection was examined in \cite{Jianq}. The authors focused on the path planning problem and proposed solutions to optimize the path taken by the UAVs-UGVs cooperative system during the detection process.
Moreover, \cite{Zhao} presented a novel approach by proposing a UAV-aided integrated platooning vehicle network for energy consumption minimization based resource management. This study emphasized the potential of using UAVs to optimize resource management in a platooning vehicle network, ultimately reducing energy consumption. 
However, none of the above works consider how to deploy UAVs so as to collect data from UGVs without the requirement of cellular connections. Previously, similar problems were addressed with fixed location users rather than moving UGVs \cite{9043712,Yunfei,Mohamed}.

In this paper, in order to overcome this limitation, we consider an integrated UAVs-UGVs network with UAVs collecting data from UGVs. This system comprises multiple UAVs and UGVs, where UGVs are dynamically associated with UAVs, and one UAV can only communicate with one UGV in each time slot. Our contributions are summarized as follows:
\raggedbottom
\begin{enumerate}
\item
	We formulate a sum-rate maximization problem
with co-optimization of UAV placement and UAVs-UGVs 	scheduling. This problem is unfortunately a mixed integer
	programming problem and thus NP-hard.
	
\item 
Nevertheless, we devise a tractable polynomial time algorithm, whose idea is to decouple it and alternately optimize the discrete and continuous variables with gradient descent and penalty-based binary relaxation, respectively. The worst-case complexity of the algorithm is also analyzed.

\item 
Finally, we provide simulation results to show the effectiveness of the proposed algorithm over two baselines. 
Video visualizations of the simulation results are available at  \textcolor{blue}{\href{https://bigbenny-tongzhang.github.io/exhibition.html}{https://bigbenny-tongzhang.github.io/exhibition.html}}.
\end{enumerate}

\section{System Model and Problem Formulation}
	\begin{figure}[t]
		\centering
		\label{Fig1}
		\includegraphics[width=0.45\textwidth]{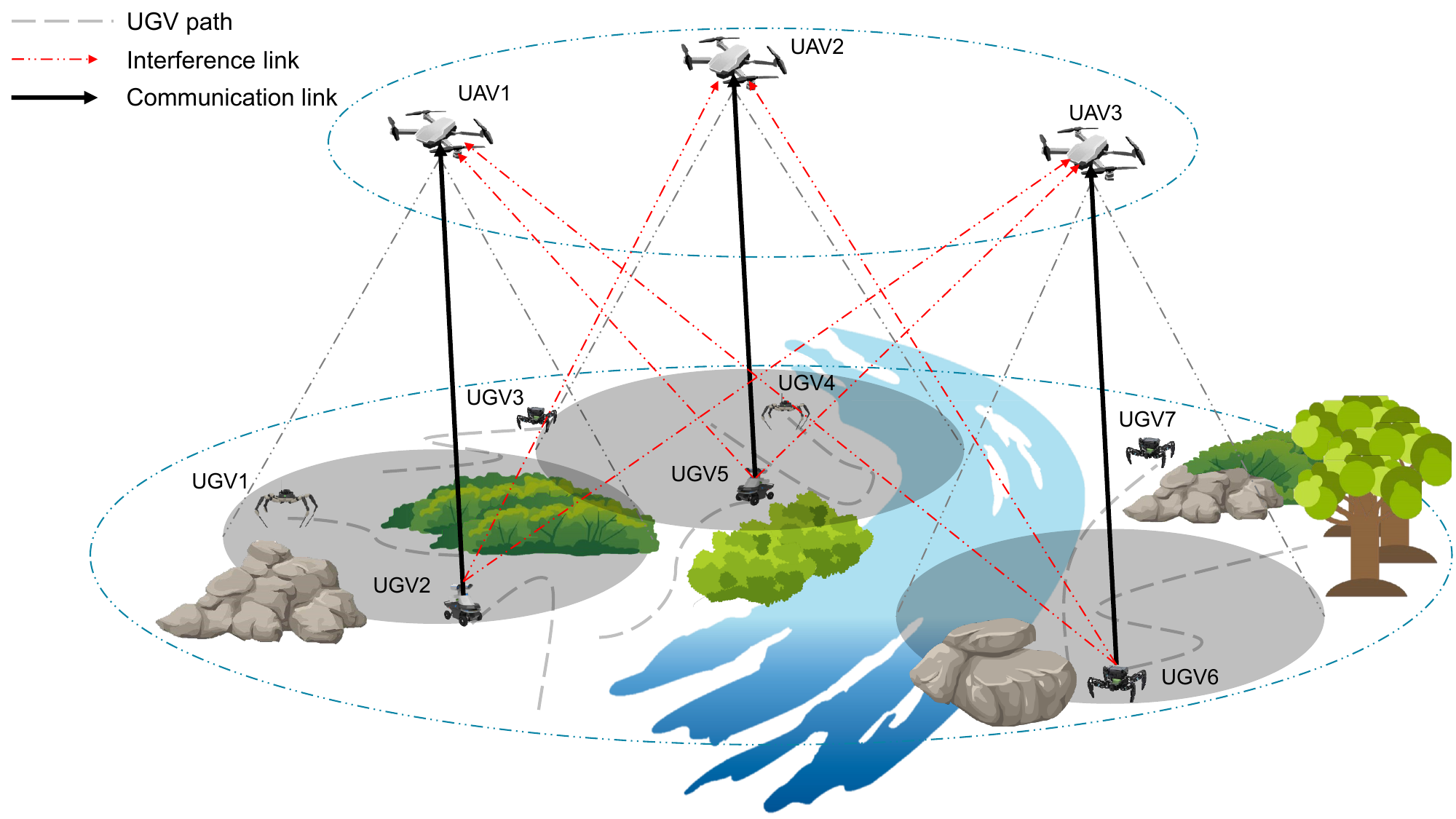}
		\caption{The considered integrated UAVs and UGVs system.}
	\end{figure}
	We consider an integrated UAVs-UGVs network in Fig.~1. In practice, this scenario is useful in supporting the flexible UGV data uploading in the wilderness without cellular connection. 
We deploy multiple UAVs to collect task data from
multiple moving UGVs. In the considered system, there are $M$ UAVs and $N$ UGVs, where both UAV and UGV are equipped with single antenna. By adopting three-dimensional Cartesian coordinate system, i.e., $(x,y,z)$, the locations of the UAV $j\in\{ 1,2,...,M\}$ and
	the UGV $i \in \{1,2,...,N\}$ are denoted by $(\bm{\alpha}_j,H_j) = (x_j^\alpha, y_j^\alpha, H_j)$ and $\bm{\beta}_i=(x_i^\beta,y_i^\beta,0)$, respectively. The communications operating in $T$ time slots, where the duration of each time slot is $\tau$. During the entire time, the UAVs are with fixed locations and the UGVs are moving with trajectory. The location of UGV $i$ at time slot $t \in \{1,2,\cdots,T\}$ is expressed as $\bm{\beta}_i[t]=(x_i^\beta[t],y_i^\beta[t],0)$.

Each UAV is allowed to communicate with  one UGV at each time slot. Mathematically, $a_{ij}\left[t\right]$ $\in$$\left\{0,1\right\}$ denotes the scheduling decision at time slot $t$, which is given by
	\begin{equation}
		a_{ij}\left[t\right]=\left\{
		\begin{aligned}			
			1&,   && \mbox{UGV $i$ is scheduled to UAV $j$ at time slot $t$},\\
			0&,   &&\mbox{else.}
		\end{aligned}
		\right. \nonumber 
    \end{equation}
	
	
The probability of having line-of-sight (LoS) links depends on the locations, heights of the UAVs and obstacles, as well as the elevation angle between  UAV and its UGV. We refer to a commonly used probabilistic path loss model provided by International Telecommunication Union (ITU-R) \cite{Abbas}. Specifically, the path loss between UAV $j$ and UGV $i$ at time slot $t$ is given according to LoS or non-line-of-sight (NLoS) 
	\begin{equation}
	  {\Lambda}_{ij}[t]=\left\{
	  \begin{aligned}
	  &\frac{4\pi^{2}d_{ij}[t]^{2}f_{c}^{2}}{c^{2}G_{t}G_{r}} \mu_{\text{NLoS}},  &&\mbox{ LoS link,}\\
	  &\frac{4\pi^{2}d_{ij}[t]^{2}f_{c}^{2}}{c^{2}G_{t}G_{r}} \mu_{\text{LoS}},  &&\mbox{ NLoS link,}&
 	  \end{aligned}
		\right.
	\end{equation}
    where $f_{c}$ is the operating frequency; $c$ is the speed of light; $G_{t}, G_{r}$ are transmitting and receiving antenna gain respectively; $\mu_{\text{NLoS}}, \mu_{\text{LoS}}$ are different attenuation factors considered for \text{NLoS} and \text{LoS}; $d_{ij}[t]=\sqrt{\lVert\bm{\alpha}_j-\bm{\beta}_i[t]\rVert+H_{j}^{2}}$ 
    is the distance between UGV $i$ and UAV $j$ at time slot $t$. The LoS link probability is given by 
    \begin{align}
    	\text{Pr}_{ij}^{\text{LoS}}[t]=\alpha\left(\frac{180}{\pi}\theta_{ij}[t]-15\right)^{\gamma},  \quad \theta_{ij}[t]\textgreater\frac{\pi}{12},
    \end{align}
	where $\theta_{ij}[t]=\sin^{-1}\left(\frac{H_{j}}{d_{ij}[t]}\right)$ is the elevation angle between UAV $j$ and UGV $i$ at time slot $t$. Note that $\alpha$ and $\gamma$ are constant values reflecting the environment impact. With the LoS probability, the NLoS probability is $\text{Pr}_{ij}^{\text{NLoS}}[t]=1-\text{Pr}_{ij}^{\text{LoS}}[t]$. When the map of environment are known in advance, we can have location-specific LoS and NLoS path loss. Otherwise, the probability model of LoS/NLoS can be deployed to generated path loss.

	Then, for UAVs-UGVs links, the received signal power of UAV $j$ from UGV $i$ at time slot $t$ is calculated as
	\begin{align}
   P_{ij}^{R}[t]=\frac{P^{T}_{i}}{ \mathbb{E}\left\{\Lambda_{ij}[t]\right\}},
	\end{align}
   where $P^{T}_{i}$ is the transmitting power of UGV $i$.	
Therefore, the SINR at UAV $j$ communicating with UGV $i$ is 
\begin{align}
	\text{SINR}_{ij}[t]=\frac{P_{ij}^{R}[t]}{I_{ij}[t]+N_0},
\end{align}
where $N_0$ denotes the power of additive White Gaussian noise (AWGN), and $I_{ij}[t]$ is the interference from other UGVs at time slot $t$, which is formulated as
	\begin{align}
			I_{ij}[t]=\sum_{p,p\neq i}^{N}\left(\sum_{q=1}^{M}a_{pq}[t]\right)P_{pj}^{R}[t],
		\end{align}
 where $\sum_{q=1}^{M}a_{pq}[t]$ $\in$ $\left\{0,1\right\}$ denotes the communication status of UGV $p$. That is, $\sum_{q=1}^{M}a_{pq}[t]=1$ meaning UGV $p$ is communicating with a UAV at transmitting power $P^{T}_{p}$. 
Based on the above model and definition, the rate between UAV $j$ and UGV $i$ at time slot $t$ can be calculated as
\begin{equation}
	\begin{aligned}
		&R_{ij}\left[t\right] = \log\left(1+a_{ij}[t] \text{SINR}_{ij}[t]\right)=\\
		&\log\left(1+\frac{a_{ij}[t]P^{T}_{i}}{\mathbb{E}\left\{\Lambda_{ij}[t]\right\}\left(\sum_{p,p\neq i}^{N}\left(\sum_{q=1}^{M}a_{pq}[t]\right)P_{pj}^{R}[t]+N_0\right)}\right),
	\end{aligned}
\end{equation}
where $\log$ refers to $\log_2$. 
	
	In total $T$ time slots, multiple UAVs are deployed at fixed locations to collect data from UGVs. 
	In order for the UAVs to collect as much data as possible, maximizing the sum rate during $T$ is set as the goal. Let us define $\mathcal{A}=\left\{a_{ij}\left[t\right], \forall i, j, t\right\}$ as the scheduling decision for all $N$ UGVs and $\mathcal{X} = \{(x_j^\alpha, y_j^\alpha, H_j),\forall j\}$ as locations for all $M$ UAVs. Our aim is to optimize $\mathcal{A}$ and $\mathcal{X}$ to maximize the achievable sum rate. Mathematically, this problem can be equivalently formulated as follows:
	\begin{subequations}
		\label{P1}
		\begin{align}
			\mathcal{P}_0:\,\,
			\max_{\mathcal{A},\mathcal{X}}~&\nonumber \sum_{t=0}^{T}\sum_{i=1}^{N}\sum_{j=1}^{M} R_{ij}\left[t\right]\\
			\textrm{s.t.}~~ 
			&\sum_{i=1}^{N}a_{ij}\left[t\right]\leq 1, \forall j,t\\
			&\sum_{j=1}^{M}a_{ij}\left[t\right]\leq 1, \forall i,t\\
			&a_{ij}[t] \in\left\{0,1\right\}, \forall i,j,t
		\end{align}
	\end{subequations}
	where 
	(7a) ensures that the UAV serves one UGV at each time slot; (7b) ensures that one UGV communicates with one UAV at each time slot;  (7c) ensures that the variables are binary. 

\section{Proposed Algorithm with Alternating Gradient Decent and Penalty-based Relaxation} 

It can be seen that Problem $\mathcal{P}_0$ is a mixed integer programming problem due to the variable types, hence it is NP-hard. Nevertheless, we develop a polynomial time algorithm with alternating gradient decent and penalty-based binary relaxation. Specifically, we decompose Problem $\mathcal{P}_0$ into two subproblems. Subproblem 1 is for  the UAVs-UGVs scheduling, and subproblem 2 is for the drone placement. These subproblems are solved alternately until convergence.

\subsection{Subproblem 1: Difference-of-Convex Penalty-based Binary Relaxation for UAVs-UGVs Scheduling}
According to penalty method, the following penalty function $\psi\left(\mathcal{A}\right)$ is added into the original objective function:
\begin{align}
	\psi\left(\mathcal{A}\right)=\eta\sum_{t=0}^{T}\sum_{i=1}^{N}\sum_{j=1}^{M}a_{ij}[t]\left(1-a_{ij}[t]\right),
\end{align}
where the binary variables $a_{ij}[t] \in\left\{0,1\right\}$ can be relaxed to continuous variables $0\leq a_{ij}[t] \leq1$ without loss of optimality if the weight parameter of penalty $\eta$ is chosen suitably. To ensure the relaxed continuous problem approximating the original binary problem, $\eta$ should be chosen in the same order as the original objective value. The relaxed continuous problem of $\mathcal{P}_0$ given drone placement is given by
\begin{subequations}
	\label{P2}
	\begin{align}
		\mathcal{P}_1:\,\,
		\max_{\mathcal{A}}~& \nonumber\sum_{t=0}^{T}\sum_{i=1}^{N}\sum_{j=1}^{M} R_{ij}\left[t\right]-\psi\left(\mathcal{A}\right)\\
		\textrm{s.t.}~~ &7(a),7(b)\\
		&0\leq a_{ij}[t] \leq1,~\forall i,j,t.
	\end{align}
\end{subequations}
Unfortunately, it is still non-convex due to concave terms in the penalty function and original object function, i.e., $-a_{ij}[t]^2$ and $-\log\left(\sum_{p,p\neq i}^{N}\left(\sum_{q=1}^{M}a_{pq}[t]\right)P_{pj}^{R}[t] + N_0\right)$.
\begin{algorithm}[t]
	\caption{Algorithm for Subproblem 1}
	\begin{algorithmic}[1]
		\State{}\textbf{Initialize}:Random initialize communicating schedule $\mathcal{A}^{0}$ and set $\text{iter}=1$, tolerance $\epsilon>0$;
		\State\textbf{Repeat}:
		\State \quad Solve Problem $\mathcal{P}_1'$ given $\mathcal{A}^{\text{iter}-1}$ and yield the optimal solution as $\mathcal{A}^{\text{iter}}$.
		\State \quad Set $\text{iter}\leftarrow \text{iter}+1$; 
		\State\textbf{Until}: The fractional increase of the objective value of $\mathcal{P}_1'$
		is below $\epsilon>0$.
	\end{algorithmic}
\end{algorithm}
Therefore, we propose to utilize the difference-of-convex procedure to convexify the concave terms. An efficient way is to implement first-order Taylor expansion on the concave terms to approximate them. For both functions, the concave function of $-a_{ij}[t]^{2}$ and $-\log\left(\sum_{p,p\neq i}^{N}\left(\sum_{q=1}^{M}a_{pq}[t]\right)P_{pj}^{R}[t] + N_0\right)$ is approximated in $(10)$ and $(11)$, respectively, where $a_{ij}[t]^{\prime}$ denotes the  scheduling decision in last iteration.
\begin{align}
	-a_{ij}[t]^{2}\leq \left(a_{ij}[t]^{\prime}\right)^{2}-2a_{ij}[t]^{\prime}a_{ij}[t]
\end{align}
\begin{figure*}
	\begin{equation}
		 	\log\left(\sum_{p,p\neq i}^{N}\left(\sum_{q=1}^{M}a_{pq}[t]\right)P_{pj}^{R}[t] + N_0\right)\geq
		 \underbrace{\log\left(\sum_{p,p\neq i}^{N}\left(\sum_{q=1}^{M}a_{pq}[t]^{\prime}\right)P_{pj}^{R}[t] + N_0\right)+	 
		 \frac{\sum_{p,p\neq i}^{N}\left(\sum_{q=1}^{M}\left(a_{pq}[t]-a_{pq}[t]^{\prime}\right)\right)P_{pj}^{R}[t]}{\sum_{p,p\neq i}^{N}\left(\sum_{q=1}^{M}a_{pq}[t]^{\prime}\right)P_{pj}^{R}[t] + N_0} }_{\text{denoted by }g_{ij}[t]}		  
	\end{equation}
\hrule
\end{figure*}
For brevity, we specify $g_{ij}[t]$ and $f_{ij}[t]$ to denote different terms in the difference-of-convex penalty-based function as shown in $(11)$ and $(12)$, respectively.
\begin{equation}
	\begin{aligned}
		&f_{ij}[t]=\log\left(a_{ij}[t]P^{T}_{i}+ N_0+\sum_{p,p\neq i}^{N}\left(\sum_{q=1}^{M}a_{pq}[t]\right)P_{pj}^{R}[t] \right)\\
	\end{aligned}
\end{equation}

Then, the $\text{iter}^{th}$-iteration convex problem in the difference-of-convex method is given by
\begin{subequations}
	\label{P3}
	\begin{align}
		\mathcal{P}_1':\,\,
		\max_{\mathcal{A}}~&\nonumber\sum_{t=0}^{T}\sum_{i=1}^{N}\sum_{j=1}^{M}\left(f_{ij}[t]-g_{ij}[t]\right)\\
		&\nonumber-\eta\sum_{t=0}^{T}\sum_{i=1}^{N}\sum_{j=1}^{M}(a_{ij}[t]+(a_{ij}[t]^{\text{iter}-1})^{2}\\
		&\nonumber-2a_{ij}[t]^{\text{iter}-1}a_{ij}[t])\\
		\textrm{s.t.}~~ &9(a),9(b). 
	\end{align}
\end{subequations}

Finally, Algorithm 1 summarizes the procedure for solving Subproblem 1. 
\subsection{Subproblem 2: Gradient Decent for Drone Placement}

Regarding the drone placement subproblem, it is worth noting that the number of variables in this subproblem is $2M$, where $M$ represents a small number of variables. Therefore, an efficient approach to solve this subproblem is by employing the gradient descent method \cite{boyd2004}. This subproblem is thus given by
\begin{subequations}
	\label{P4}
	\begin{align}
		\mathcal{P}_2:\,\,
		\max_{\mathcal{X}}~&\nonumber \sum_{t=0}^{T}\sum_{i=1}^{N}\sum_{j=1}^{M} R_{ij}\left[t\right]\\
		\textrm{s.t.}~~&x_j^\alpha\geq0, y_j^\alpha\geq0,~\forall j.
	\end{align}
\end{subequations}
\begin{algorithm}[t]
	\caption{Overall Algorithm for  Problem $\mathcal{P}_0$}
	\begin{algorithmic}[1]
		\State{}\textbf{Initialize}:Initialize drone placement $\mathcal{X}^{0}$ and set $r=1$, tolerance $\epsilon>0$;
		\State\textbf{Repeat}:
		\State \quad Solve Problem $\mathcal{P}_1$ by Algorithm $1$ for given $\mathcal{X}^{r-1}$  and denote the optimal solution as $\mathcal{A}^{r}$
		\State \quad Solve Problem $\mathcal{P}_2$ by gradient decent for given $\left\{\mathcal{A}^{r},\mathcal{X}^{r-1}\right\}$ and denote the optimal solution as $\mathcal{X}^{r}$
		\State \quad Set $r\leftarrow r+1$;
		\State\textbf{Until}: The fractional increase of the objective value of $\mathcal{P}_0$
		is below $\epsilon>0$.
	\end{algorithmic}
\end{algorithm}
\subsection{Overall Algorithm and Complexity Analysis}
Algorithm 2 summarizes the  proposed algorithm with gradient descent and penalty-based ralexation. 
The convergence of the overall algorithm is guaranteed by \cite{boyd2004, de2020sequential}. Furthermore, the worst-case computational complexity of Algorithm 2 is given by the following proposition:

$\underline{\text{Proposition 1}}$: The worst-case computational complexity of Algorithm 2 is $\mathcal{O}$$\left(K_{1}\left(K_{2}(MNT)^{3.5}+MCK_{3}\right)\right)$, where $K_{1}$, $K_{2}$ and $K_{3}$ are the number of outer, difference-of-convex and gradient-descent iterations, respectively and $C$ is the derivative computational cost of the objective function.

\begin{IEEEproof}
	Since Problem $\mathcal{P}_1'$ is convex, the worst-case computational complexity of solving Problem $\mathcal{P}_1'$ using interior point method is $\mathcal{O}\left((MNT)^{3.5}\right)$. Then for the difference of convex procedure, the convergence requires $K_{2}$ iterations. Secondly, the computational complexity of solving Problem $\mathcal{P}_2$ using gradient descent method is $\mathcal{O}\left(MCK_{3}\right)$, where $K_{3}$ represents the number of iterations required to meet the desired criteria, and $C$ denotes the derivative computational cost of the objective function. And the convergence of alternating optimization requires $K_{1}$ rounds, where $K_{1}$ is a finite number and not very large in practice. Finally, the computational complexity of Algorithm 2 is $\mathcal{O}$$\left(K_{1}\left(K_{2}(MNT)^{3.5}+MCK_{3}\right)\right)$.
\end{IEEEproof}

\begin{figure*}[t]
	\centering
	\renewcommand{\thefigure}{2}
	\begin{minipage}[!h]{\linewidth}
		\subcaptionbox{Time slot:$2$}{\includegraphics[width=.245\linewidth]{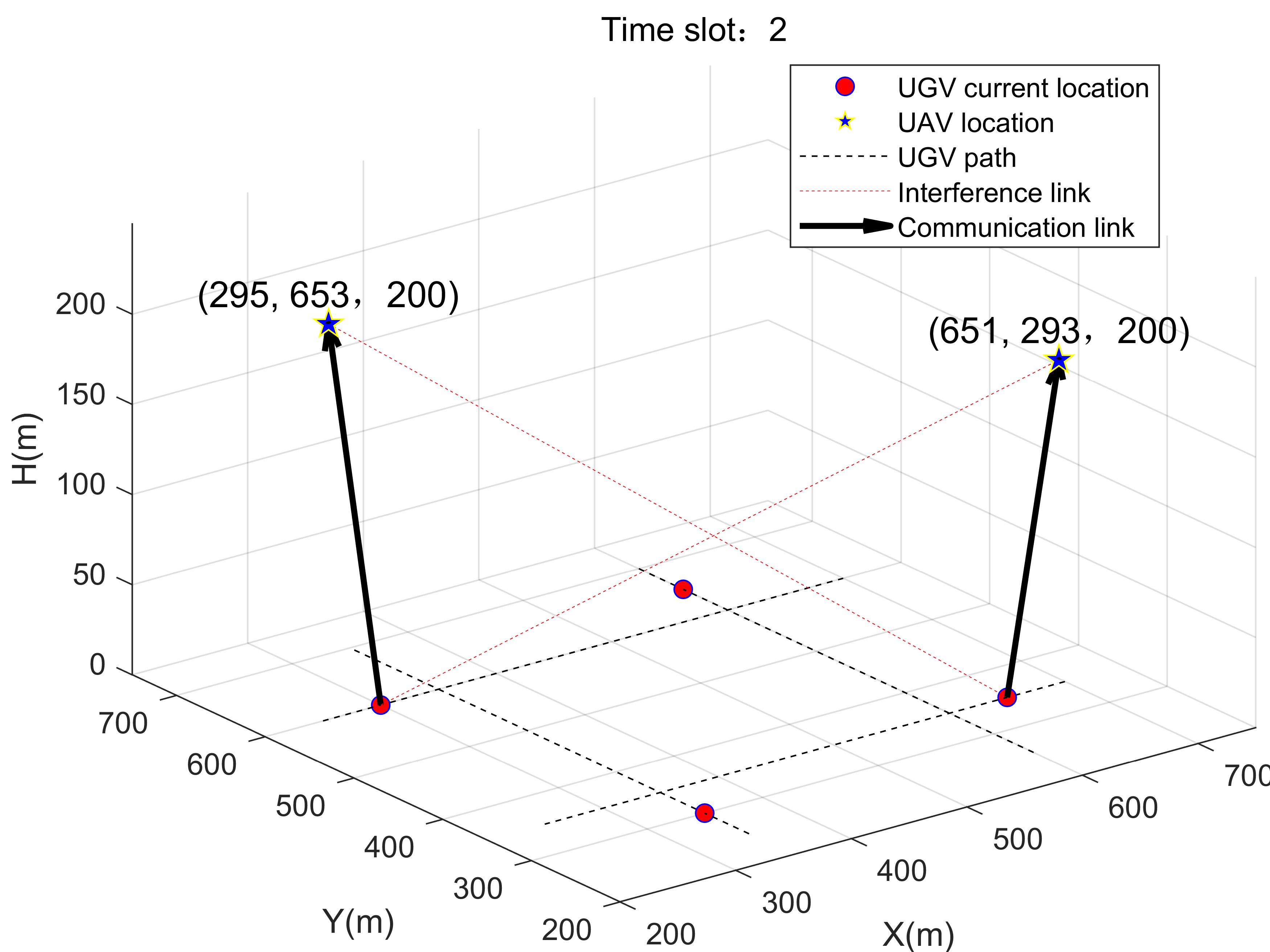}}
		\subcaptionbox{Time slot:$5$}{\includegraphics[width=.245\linewidth]{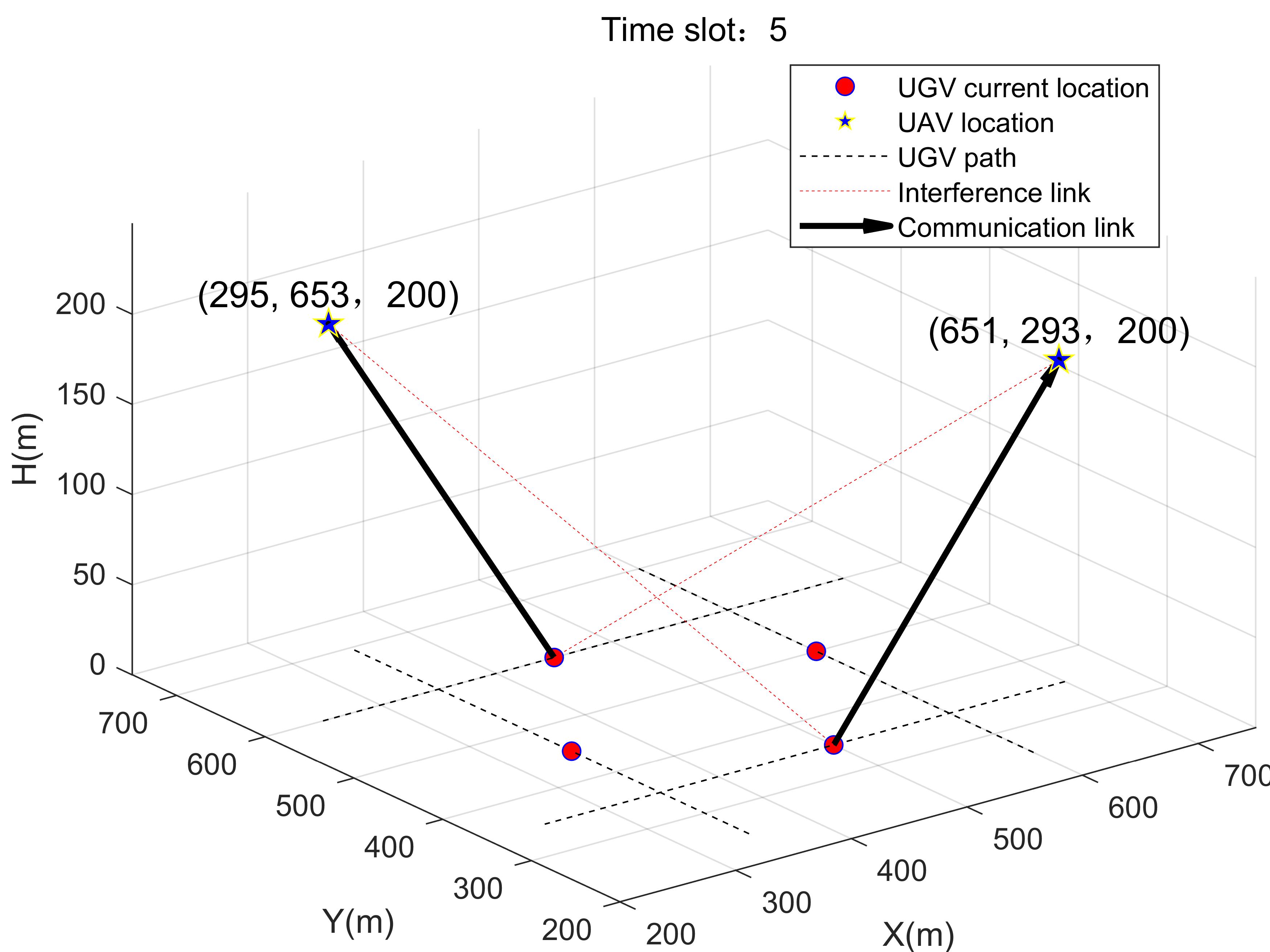}}
		\subcaptionbox{Time slot:$6$}{\includegraphics[width=.245\linewidth]{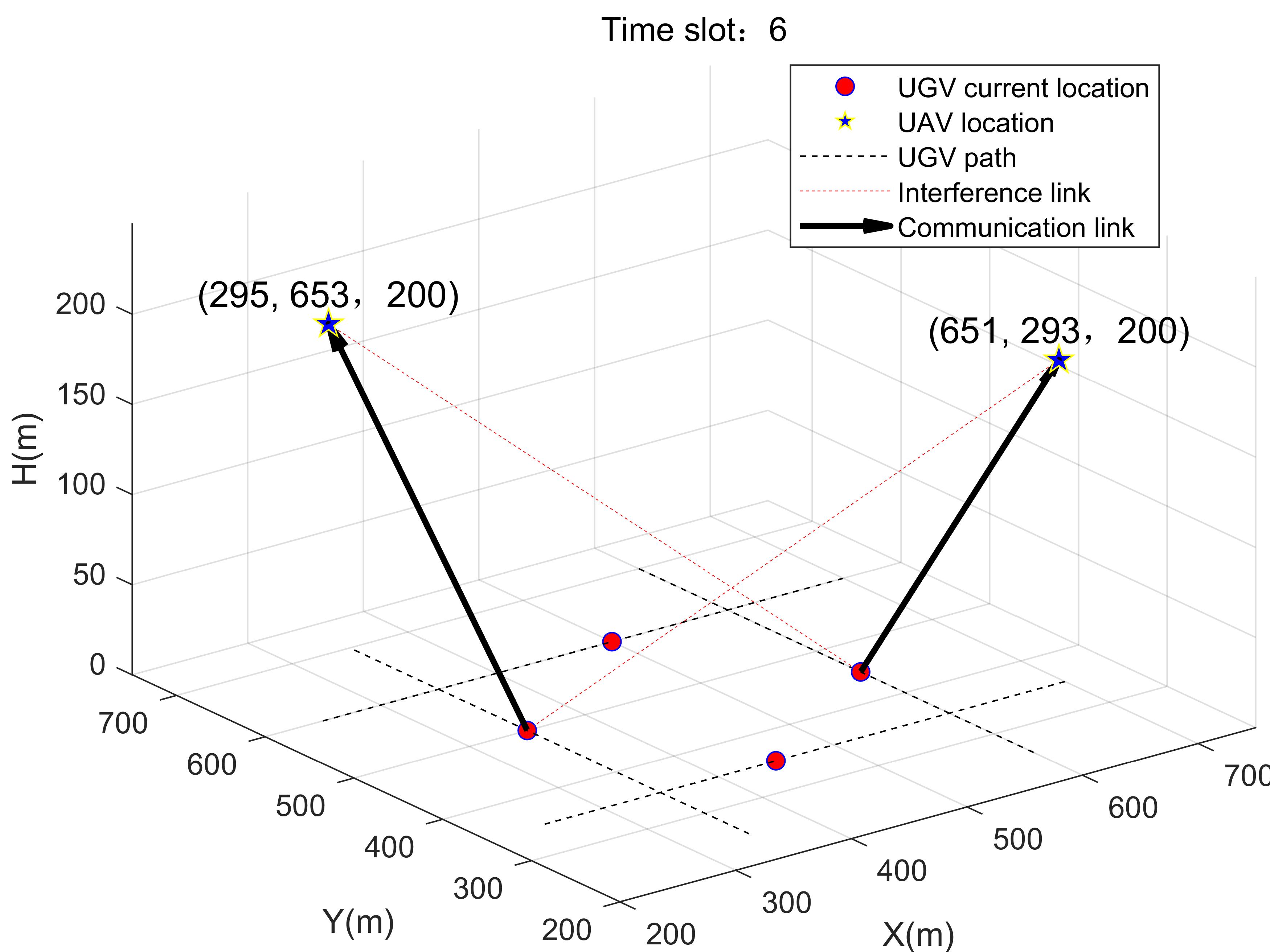}}
		\subcaptionbox{Time slot:$9$}{\includegraphics[width=.245\linewidth]{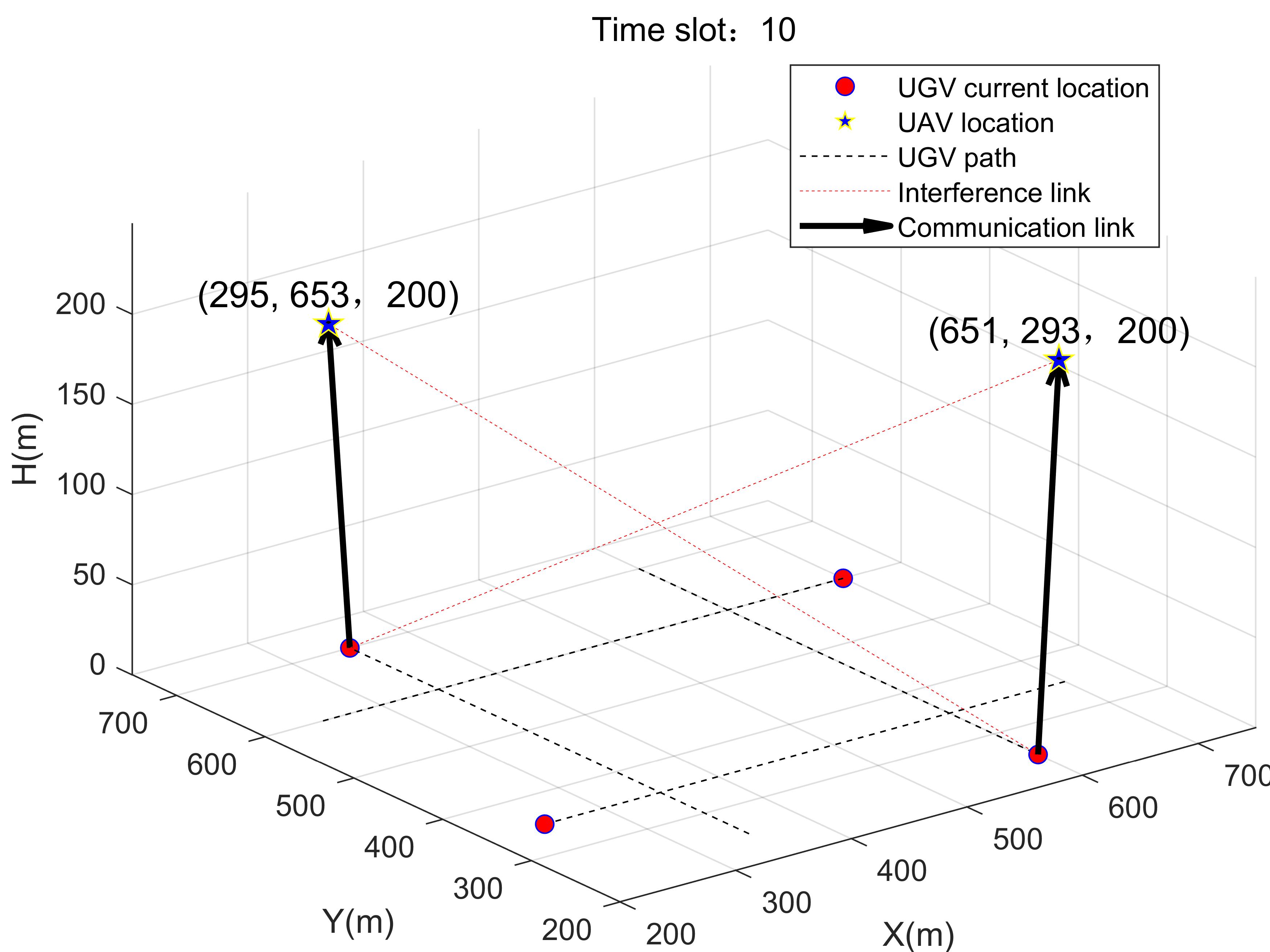}}
		\caption{Communication and interference links at different time  slots in line trajectory system}\label{Links1}
	\end{minipage}
\end{figure*}
\begin{figure*}[t]
	\centering
	\renewcommand{\thefigure}{3}
	\begin{minipage}[!h]{\linewidth}
		\subcaptionbox{Time slot:$2$}{\includegraphics[width=.245\linewidth]{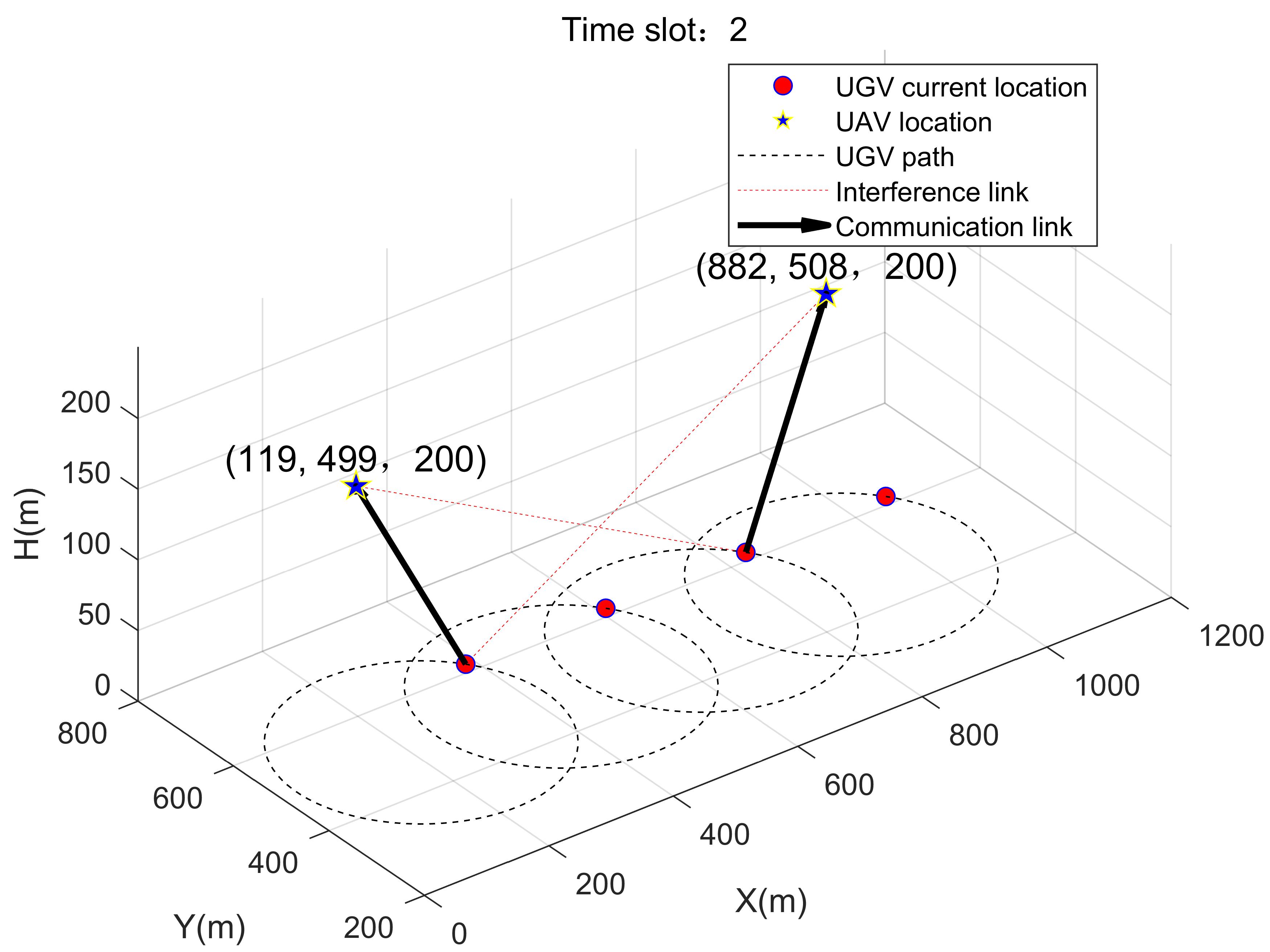}}
		\subcaptionbox{Time slot:$5$}{\includegraphics[width=.245\linewidth]{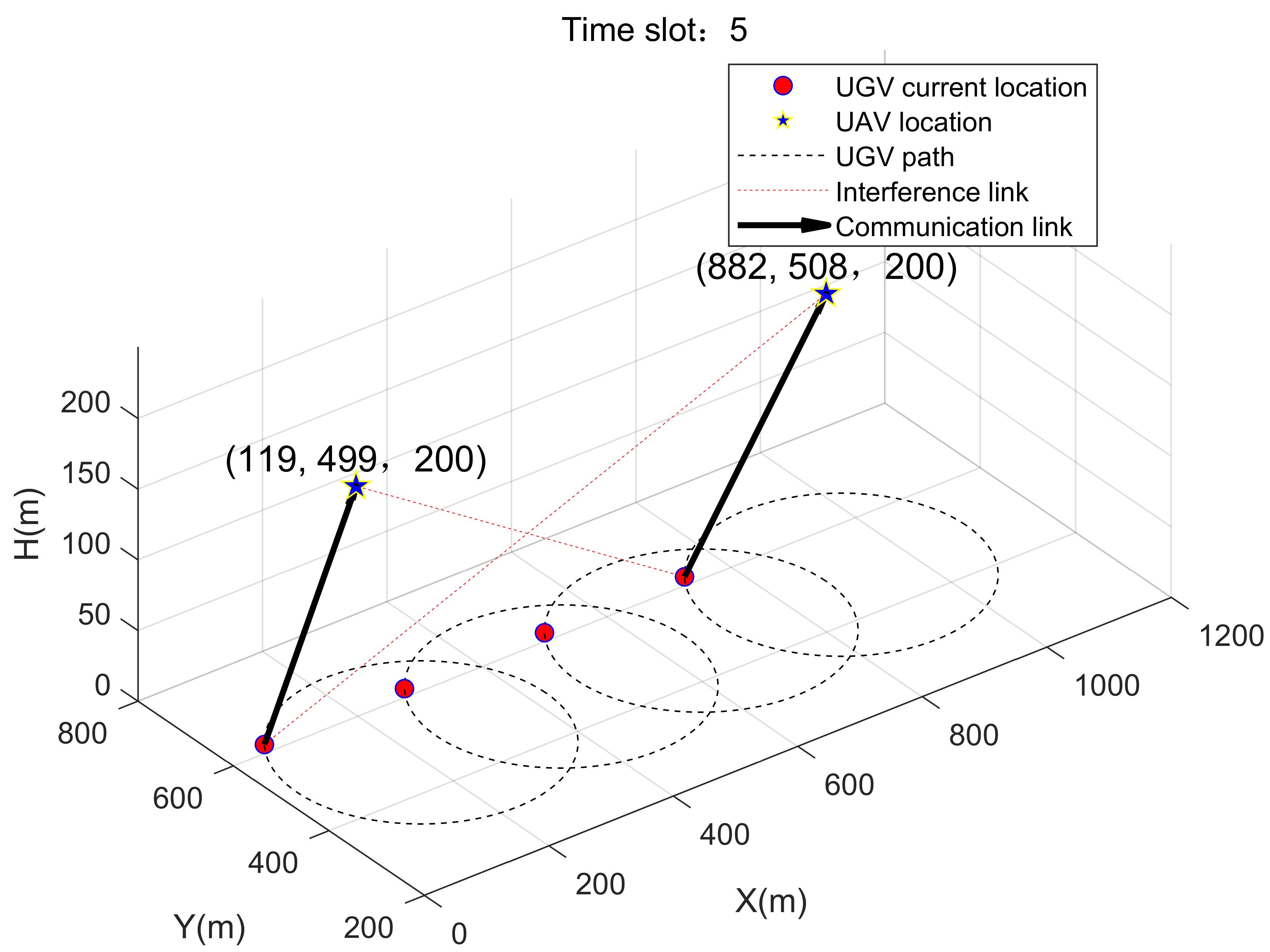}}
		\subcaptionbox{Time slot:$7$}{\includegraphics[width=.245\linewidth]{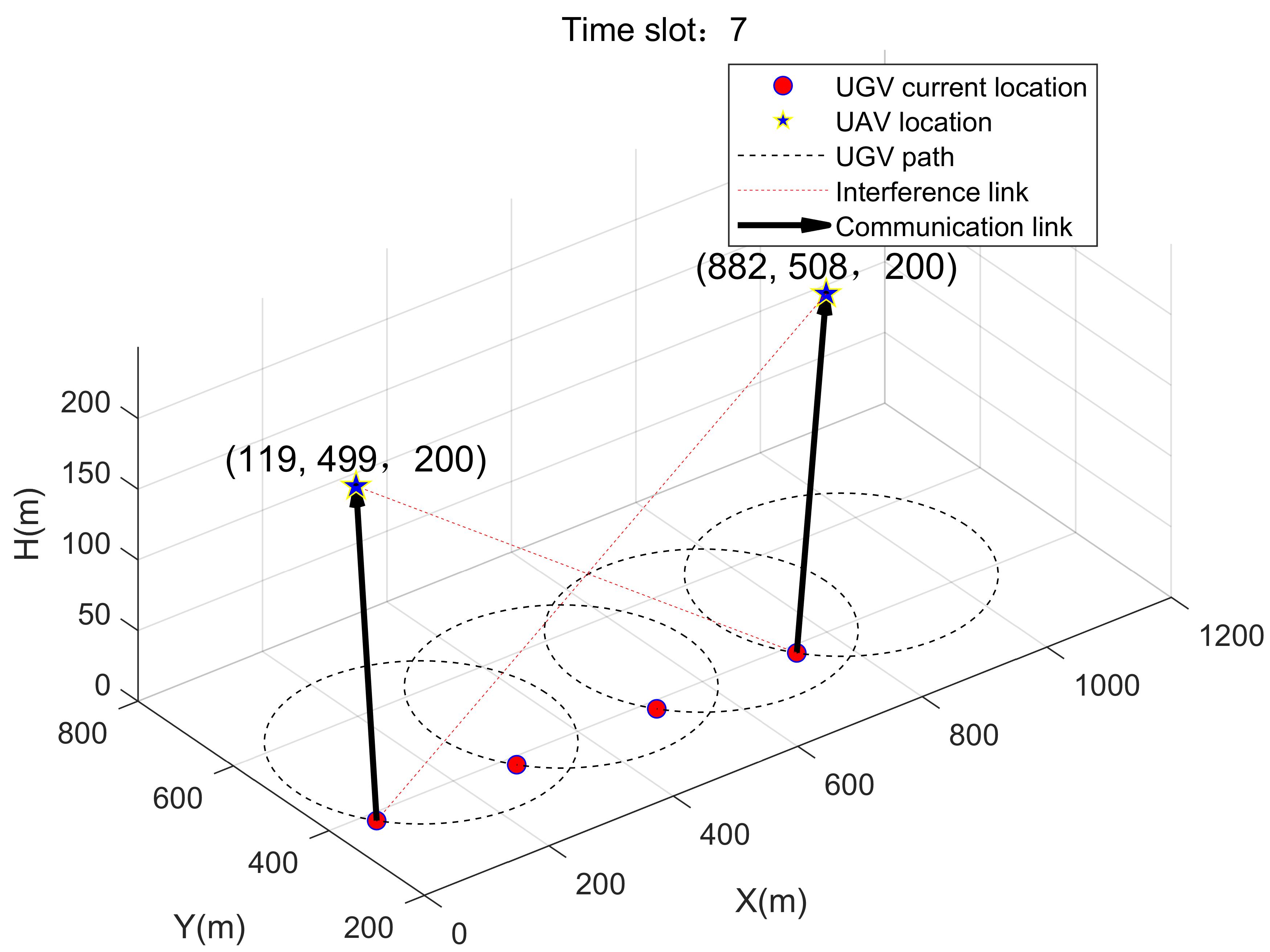}}
		\subcaptionbox{Time slot:$9$}{\includegraphics[width=.245\linewidth]{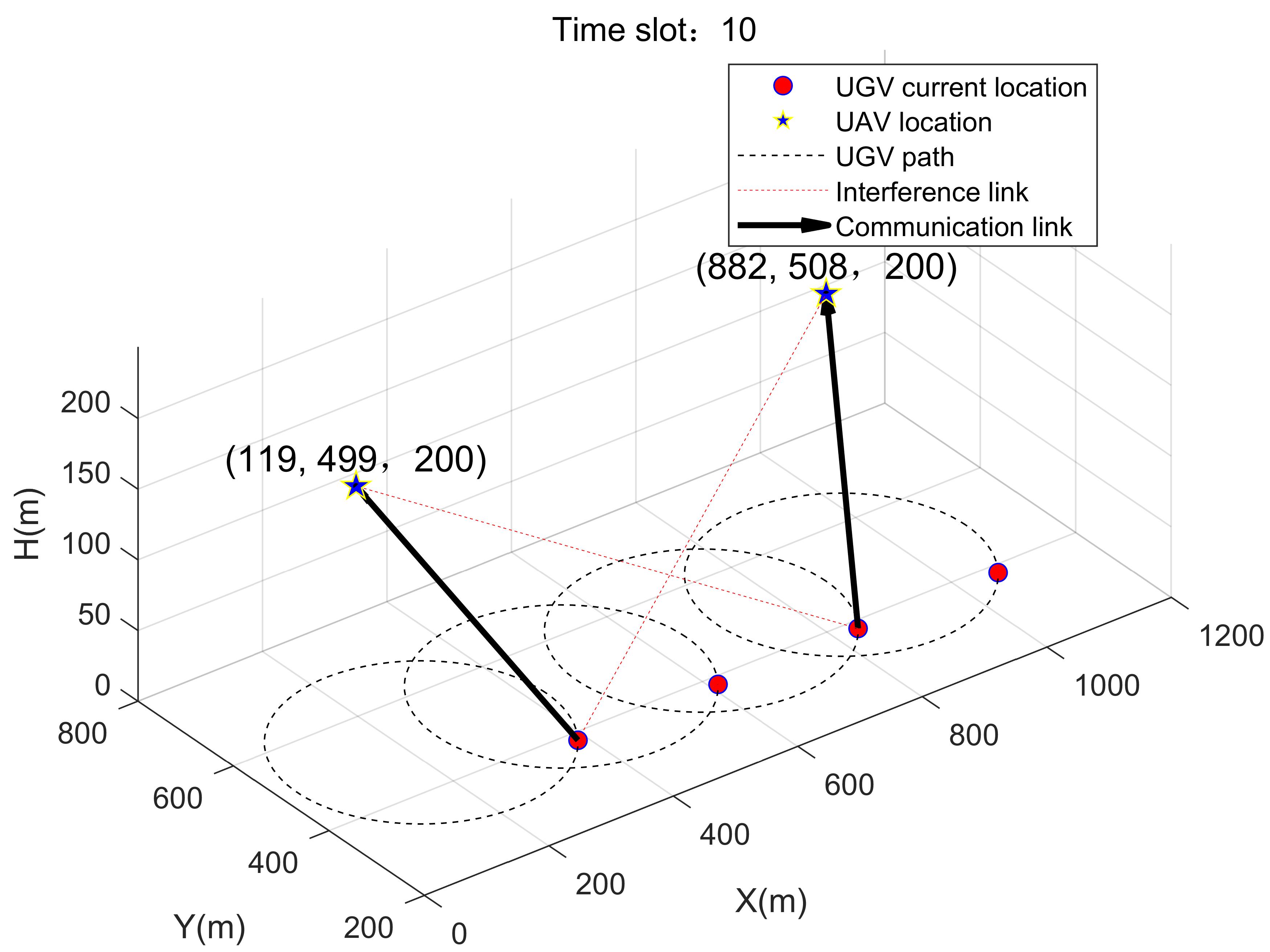}}
		\caption{Communication and interference links at different time slots in circle trajectory system}\label{Links}
	\end{minipage}
	\vspace{-0.3cm}
    \end{figure*}
\section{Simulation Results}

In this section, simulation results are provided to verify our proposed algorithm. We consider two scenarios, one of which has a UGV trajectory consisting of intersecting lines as depicted in Fig. 2, while the other system consists of
 overlapping circles in Fig. 3. Simulation parameters of the two systems are listed in table~1. 
Before analyzing the simulation results, we first introduce the obbreviations for two baselines, listed as follows:
\begin{itemize}
	\item[$\bullet$]\textbf{Fixed selection (Baseline 1)} selects the two UGVs with the maximum distance between them   during the entire process and then optimizes  UAV positions.
	\item[$\bullet$]\textbf{Random selection (Baseline 2)} randomly chooses two UGVs during the entire process and then optimizes the UAV positions.
\end{itemize}

In the line trajectory system, we set the UGV trajectory length to 450m. Partial results are shown in Figs. 2(a)-(d). UAVs are strategically deployed diagonally to the UGV trajectory to mitigate interference, which significantly affects the achievable sum rate in our system. Moreover, Figs. 2(b) and Figs. 2(c) highlight a noticeable transition between time slot 5 and 6, where UGVs previously communicating become farther from UAVs compared to the other two UGVs. This transition leverages the increased distance to smoothly switch communications from the initial UGVs to the remaining ones.
\begin{table}[t]
	\renewcommand{\arraystretch}{1.5}   
	\begin{center}   
		\caption{Simulation Parameters}  
		\label{table:1} 
		\begin{tabular}{|m{1.6cm}<{\centering}|m{2.9cm}<{\centering}|m{1.3cm}<{\centering}|m{1.3cm}<{\centering}|}   
			\hline    \textbf{Variables} & \textbf{Interpretation} & \textbf{Line Trajectory} & \textbf{Circle Trajectory} \\  
			\hline    $M$ &UAV number  & 2  & 2   \\ 
			\hline    $N$ &UGV number & 4  & 4  \\ 
			\hline   \textbf{$T$} &Number of time slots& 10 & 10  \\  
			\hline    ${f}_{c}$ &Operating frequency& 2 MHz & 2 MHz  \\ 
			\hline   \textbf{$P_{i}^{T}, \forall i$} &Transmit power& 1 W & 1 W  \\
			\hline   \textbf{$H_{j}, \forall j$} &UAV deployment height& 200 m & 200 m  \\ 
			\hline   \textbf{$N_{0}$} &Power of AWGN& -90dB  & -90dB  \\ 
			\hline   
		\end{tabular}   
	\end{center}   
\end{table}
\begin{figure}[t]
	\centering
	\label{Fig10}
	\includegraphics[width=0.25\textwidth]{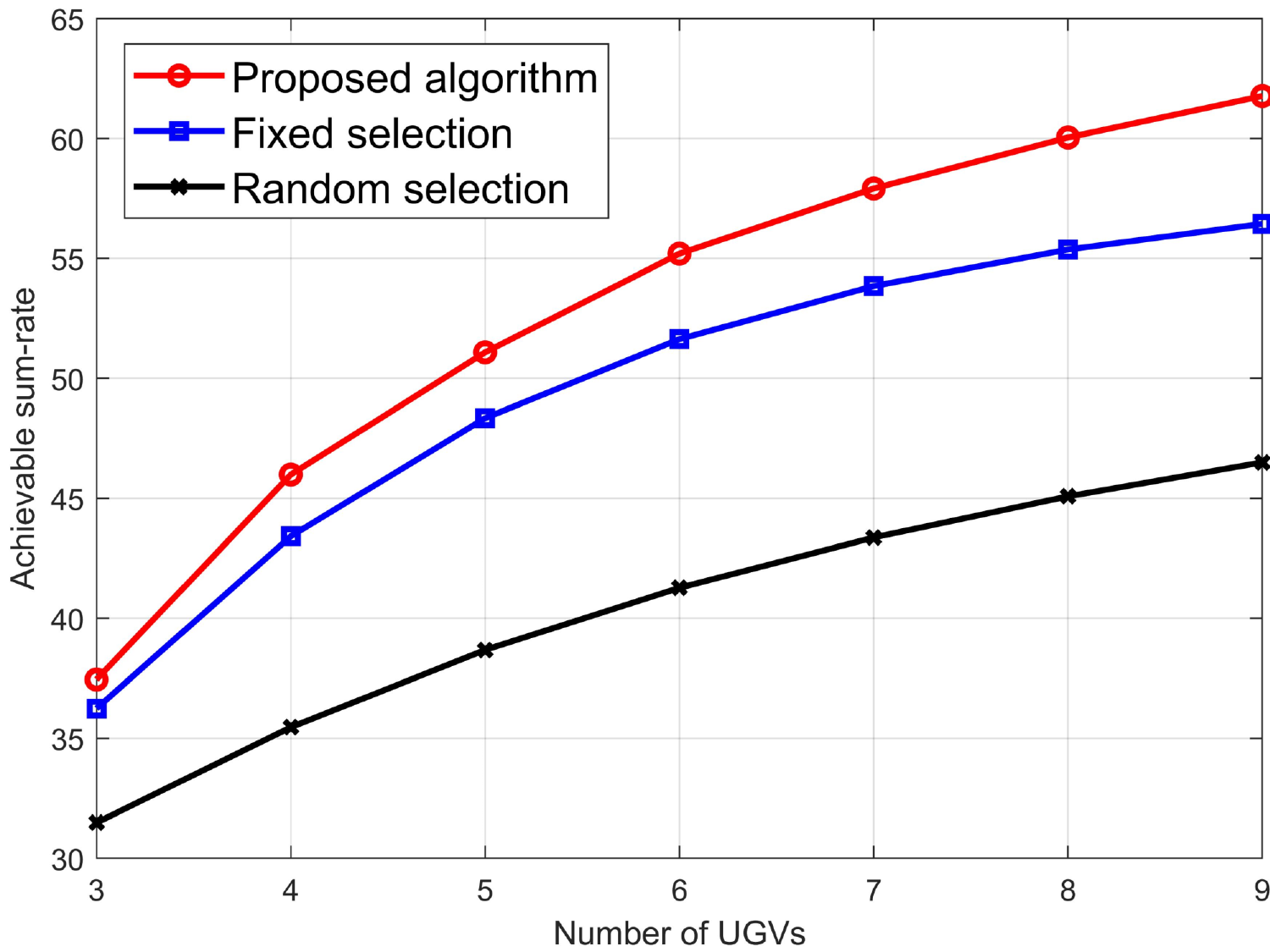}
	\caption{Number of UGVs v.s. achievable sum-rate.}
\end{figure}
In the circle trajectory system, we set the UGV trajectories with a radius of $200m$. Partial results can be found in Figs.~3(a)-(d).  
As is shown, the right UAV is deployed at the center of the intersection of the right two trajectories to facilitate the scheduling of the right two UGVs. Due to interference, the left UAV is not symmetrically deployed to the right UAV's placement.
Instead, it is deployed slightly to the left of the center of the first left UGV trajectory to mitigate the interference it experiences. As a result, the right UGV communicates with the right two UGVs and schedules the one closer to itself; the left UGV only communicates with the first UGV on the left side. 

In Fig. 4, we investigate the performance of the proposed algorithm in circle system, under varying numbers of UGVs. It can be observed that our algorithm consistently outperforms the other two algorithms.
 Moreover, the performance difference between our algorithm and Fixed selection increases as the number of UGVs grows. This can be attributed to the circle trajectory system we used, where with a smaller number of UGVs, the dominant factor is the strong interference due to the closer proximity so that Fixed selection can effectively reduce the impact of interference but still falls slightly short in performance compared to our algorithm. As the number of UGVs increases, certain UGVs become far from each other so that the influence of interference gradually diminishes. Fixed selection misses opportunities for better communication scheduling while our algorithm yields superior communication scheduling solutions and thus exhibits much better performance, which also explains why our algorithm demonstrates better performance compared to Random selection  that can only offer random communication scheduling strategies.

\section{Conclusion}
We considered an integrated UAVs-UGVs network, where UAV placement and UAVs-UGVs link scheduling are approximately optimized by an alternating algorithm.
The effectiveness of proposed algorithm was demonstrated. In  the future, more complex scenarios can be considered.

\bibliographystyle{IEEEtran}
\bibliography{ICCT}
\end{document}